\documentclass[twoside,superscriptaddress,twocolumn,floatfix,a4paper,pra]{revtex4}

\usepackage{color}
\usepackage{graphicx}
\usepackage[utf8x]{inputenc}
\usepackage[T1]{fontenc}
\usepackage{siunitx}
\usepackage{amstext}
\usepackage{textgreek}
\usepackage{hyperref}
\usepackage{epstopdf}
\usepackage{amsfonts}
\usepackage{amsmath}
\usepackage{multirow}

\usepackage{marvosym}
\usepackage[rgb]{xcolor}

\begin{document}

\title{Implementation of an efficient linear-optical quantum router}

\author{Karol Bartkiewicz} \email{bark@amu.edu.pl}
\affiliation{Faculty of Physics, Adam Mickiewicz University,
PL-61-614 Pozna\'n, Poland}
\affiliation{RCPTM, Joint Laboratory of Optics of Palacký University and Institute of Physics of Czech Academy of Sciences, 17. listopadu 12, 771 46 Olomouc, Czech Republic}

\author{Antonín Černoch} \email{acernoch@fzu.cz}
\affiliation{Institute of Physics of Czech Academy of Sciences, Joint Laboratory of Optics of PU and IP AS CR, 17. listopadu 50A, 772 07 Olomouc, Czech Republic}
   
\author{Karel Lemr}
\email{k.lemr@upol.cz}
\affiliation{RCPTM, Joint Laboratory of Optics of Palacký University and Institute of Physics of Czech Academy of Sciences, 17. listopadu 12, 771 46 Olomouc, Czech Republic}   

\begin{abstract}
In this paper, we report on experimental implementation of a linear-optical quantum router. Our device allows single-photon polarization-encoded qubits to be routed coherently into two spatial output modes depending on the state of two identical control qubits. The polarization qubit state of the routed photon is maintained during the routing operation. The success probability of our scheme can be increased up to 25\% making it the most efficient linear-optical quantum router known to this date.
\end{abstract}

\date{\today}

\maketitle
%\section{Introduction}
For several decades, scientists have been aware of significant benefits allowing quantum information processing technologies to surpass their classical counterparts \cite{kniha1,kniha2}. Recent technological development allows these benefits to be tested experimentally and in some cases also implemented in practical devices. Quantum cryptography serves as typical examples of theoretical ideas which successfully evolved into user-ready industrial applications~\cite{Georgescu12}. Over the recent years, a number of research groups all around the world have implemented experimental quantum-optical networks to test quantum information protocols in more or less practical conditions \cite{143km,rele}. % for example \cite{bib:Sipahigil847}. 

\begin{figure}
\includegraphics[scale=0.85]{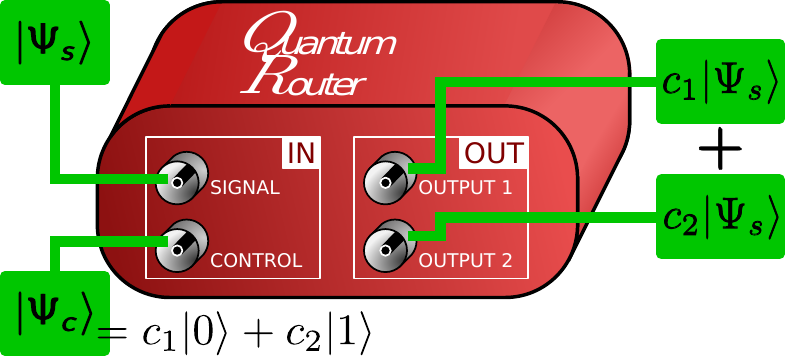}
\caption{\label{fig:concept}Conceptual scheme of a quantum router. Signal qubit is coherently routed into a superposition of output spatial modes with amplitudes given by the state of the control qubit.}
\end{figure}

So far the majority of experimental quantum networks was limited to peer-to-peer communications between two parties. Practical implementation of quantum communications networks, however, needs to address the problem of scalability to serve large numbers of users. Similarly to classical computer networks, their quantum analogues would require routing protocols to direct the signal from its source to destination \cite{routovani}. A conceptual scheme of a quantum router is depicted in Fig.~\ref{fig:concept}. The signal and control qubits denoted $|\psi_s\rangle$ and $|\psi_c\rangle = c_1|0\rangle + c_2|1\rangle$, respectively, serves as the router input. Based on the state of the control qubit, the signal is coherently forwarded to two output ports. Thus, the transformed signal state reads
\begin{equation}
|\psi_s\rangle \rightarrow c_1 |\psi_s\rangle_\mathrm{OUT1} + c_2 |\psi_s\rangle_\mathrm{OUT2},
\end{equation}
where indices OUT1 and OUT2 denote the two output ports. The quantum routing transformation belongs to a broader class of quantum state fusion protocols with the specific requirement to use spatially separate output ports \cite{Vitelli:fusion}. 
%\xxx{(should we cite something here?)}.

Quantum routers have been investigated both theoretically and experimentally for various experimental platforms~\cite{lemr:router1,Hall11,Yuan:router,Chen:router,Aoki09,Hoi11,
Zhou13,Zueco09,Lemr:router2,Lu14,Zhao14,Bartk14,Yan14,Sazim15,Chen16,Li16,Cao17,Gu17}. Not all of these implementations can, however, be considered as fully quantum. In some cases, the routing information is classical and thus the router only semi-quantum \cite{Hall11,Yuan:router}. Other implementations rely on non-linear interaction \cite{Chen:router} or combine various non-optical physical platforms making them impractical for realistic quantum networks \cite{Aoki09,Hoi11,Zhou13}. There are implementations that disturb the inserted signal state and thus can not even be considered quantum routers at all \cite{Chang12}. A general quantum state fusion protocol implemented in Ref.~\cite{Vitelli:fusion} meets all the requirements for a quantum router, but operates with a rather low success probability of 1/8 (assuming feed-forward corrections).

In this paper, we report on an experimental implementation of a linear-optical quantum router based on our original theoretical proposal \cite{lemr:router1}. In contrast to the previous implementations, our device can reach success probability (routing efficiency) of up to 1/4. To our best knowledge, this makes it the most efficient quantum router on the platform of linear optics. 
Our device manages to reach such success probability by using of two identical copies (up to a constant phase shift) of the control qubit to route one signal qubit. Unless the control qubit is obtained from a computationally difficult operation, preparation of two of them does not represent a serious obstacle to the practical usage of our routing protocol. Note that the linear-optical implementation of quantum state fusion~\cite{Vitelli:fusion} also requires a third photon to be used as an ancilla.

%\section{Router construction}
\begin{figure}
\includegraphics[scale=0.78]{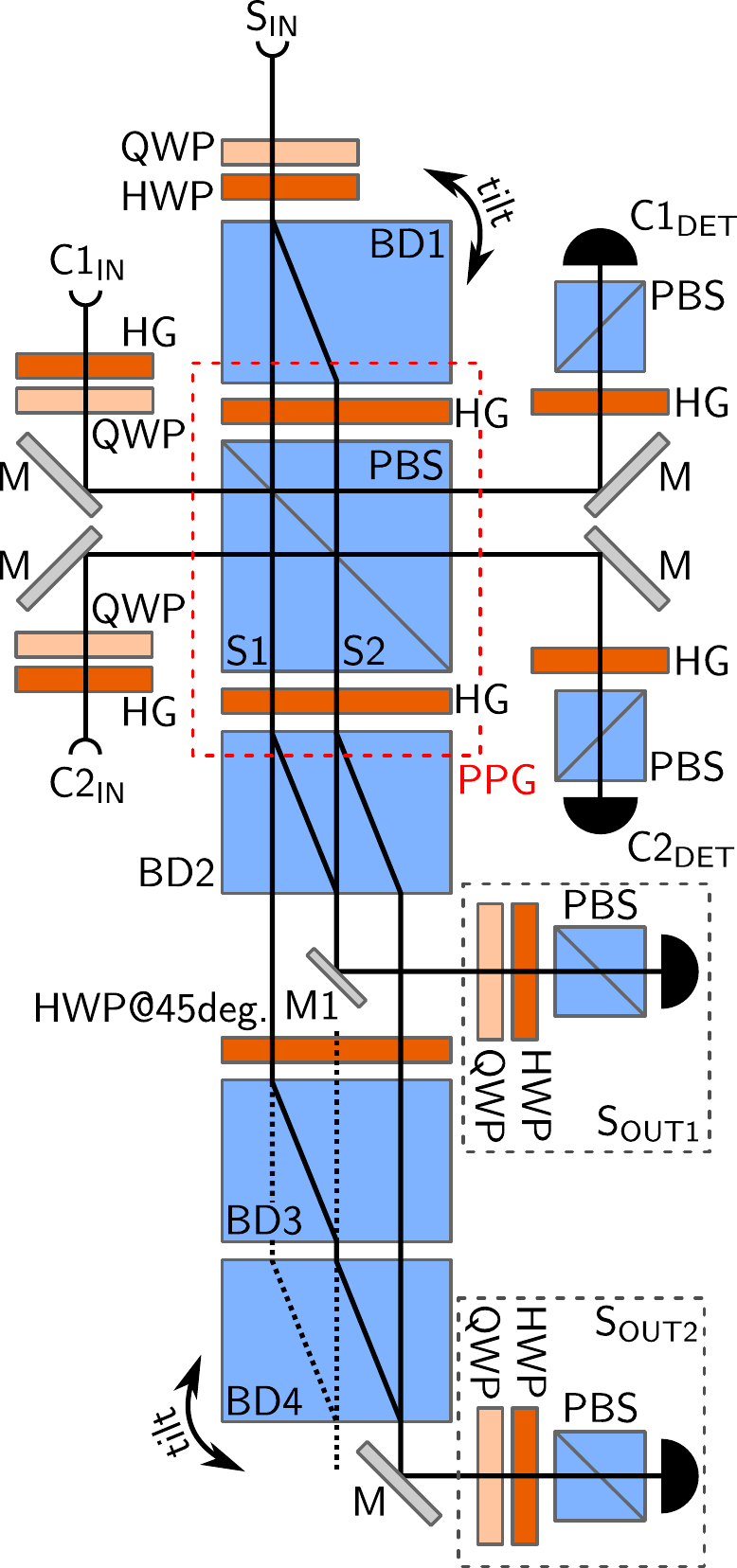}
\caption{\label{fig:setup}Experimental setup implementing the quantum router. Signal and control photons are inserted at S$_\mathrm{IN}$, C1$_\mathrm{IN}$ and C2$_\mathrm{IN}$, respectively. Control qubits are detected after being projected onto horizontal polarization states by detectors C1$_\mathrm{DET}$ and C2$_\mathrm{DET}$. Signal output state leaves the setup by two output ports denoted S$_\mathrm{OUT1}$ and S$_\mathrm{OUT2}$, where polarization analysis and detection of the signal takes place. Individual components are labeled as follows: PBS -- polarizing beam splitter, BD -- beam divider, HWP (QWP) -- half- (quarter-)wave plate, HG -- Hadamard gate (HWP rotated by 22.5 deg. with respect to horizontal polarization direction), M -- mirror. Under normal operation, beams propagate along solid black lines. For coherence testing (as explained in the text), beams trajectories are changed to black dotted lines by removing mirror M1 and beam displacer BD3.}
\end{figure}
\emph{Router construction} --- In this experiment, we used a typical three-photon source based on spontaneous parametric down-conversion (SPDC) and attenuated coherent state. A femtosecond laser pulse spectrally centered at \SI{826}{\nano\meter} with width of \SI{10}{\nano\meter} and \SI{80}{\mega\hertz} repetition rate is subjected to a second-harmonics generation unit converting its central wavelength to \SI{413}{\nano\meter}. The unconverted signal at the fundamental wavelength is attenuated and serves to prepare the signal photon. The frequency-doubled beam is spatially filtered and used to pump two BBO crystals that generate photon pairs (control photons) in a type-I SPDC process. Power of the pumping beam at the crystal is about \SI{100}{\milli\watt}. More details on the three-photon source are given in the Supplementary Material.

The working principle of our device can be understood by analyzing the experimental setup depicted in Fig.~\ref{fig:setup}. The signal and control qubits are encoded in polarizations of single photons.
Logical qubit states $|0\rangle$ and $|1\rangle$ are associated with the horizontal $|H\rangle$ and vertical $|V\rangle$ polarization of the photons respectively. This specific experimental construction has been selected with the interferometric stability in mind. For that reason, the router is based on four beam dividers forming a complex but highly stable interferometer.
The signal qubit $|\psi_s\rangle = \alpha |H\rangle + \beta |V\rangle$ is prepared at the $S_\mathrm{IN}$ port by quarter and half wave plate. Than it enters the first beam divider where its horizontal and vertical components are split into the spatial modes with the amplitudes $\alpha$ and $\beta$. Both these modes are subjected to a Hadamard gate implemented by a half-wave plate rotated by 22.5 deg. with respect to the horizontal polarization orientation. Subsequently, the two signal modes impinge on a polarizing beam splitter. On this beam splitter, each of these modes interacts with one of the two control qubits that have been prepared in the state $|\psi_c\rangle = \frac{1}{\sqrt{2}}\left(|H\rangle + \mathrm{e}^{i \varphi}|V\rangle\right)$. 
Note that the second control qubit simultaneously undergoes a transformation $\varphi \rightarrow \varphi + \pi$. This can be achieved either by a HWP or as in our case directly during the state preparation.
The parameter $\varphi$ is the real-valued parameter defining the routing amplitudes in the output ports. After interacting with the signal mode, each of the control qubits undergoes a Hadamard transform (using a half-wave plate) and is subsequently projected onto horizontal polarization and detected. The block of half-wave plates and the polarizing beam splitter together implement the programmable phase gate (PPG) on each of the signal modes \cite{micuda}. As a result, the signal modes acquire the phase shifts $\varphi$ and $\varphi + \pi$. Thus, the signal photon state can be expressed in the form of
\begin{equation}
|\psi_s\rangle = \frac{\alpha}{\sqrt{2}} \left(|H\rangle +  \mathrm{e}^{i\varphi}|V\rangle \right)_\mathrm{S1} + \frac{\beta}{\sqrt{2}} \left(|H\rangle -  \mathrm{e}^{i\varphi}|V\rangle \right)_\mathrm{S2},
\end{equation}
where indices S1 and S2 denote the spatial signal modes. These signal modes then undergo another Hadamard gate. In the final step, both these spatial signal modes are recombined on additional beam dividers. As a result the output signal state reads
\begin{equation}
\label{eq:final}
\begin{matrix}
|\psi_s\rangle & = & \cos\tfrac{\varphi}{2}\left( \alpha |H\rangle + \beta |V\rangle\right)_\mathrm{OUT1} +\\
&-& i\sin\tfrac{\varphi}{2}\left( \alpha |H\rangle + \beta |V\rangle\right)_\mathrm{OUT2}.
\end{matrix}
\end{equation}
The phase shift $\pi/2$ between the modes is insignificant and can be corrected by a phase shifter.

Projecting both the control qubits solely onto horizontal polarization makes both the included PPG gates operate with success probability of 1/4, thus the router performs with the success probability of 1/16. One can immediately increase the success probability twice by post-selecting also onto vertical polarization projection on the control qubits. If both of them are simultaneously projected onto vertical polarization, the router transformation remains identical but the output modes have to be classically swapped (e.g. using a classical optical switch). Yet another improvement in the success probability can be reached if a feed-forward correction is implemented. As it was presented in Ref.~\cite{Lemr:router2}, by means of a feed-forward correction, the PPGs can operate with success probability of 1/2, the router would thus reach the success rate of 1/4.

In the experiment, we post-select the successful router operation on three-fold coincidence detections. The two possible valid three-fold detections are the coincident detection in both control qubit output modes (detectors C1$_\mathrm{DET}$ and C2$_\mathrm{DET}$) together with either detection in the first signal output port S1$_\mathrm{OUT}$ or the second signal output port S2$_\mathrm{OUT}$. We refer to these three-fold coincident detections as CC1 and CC2, respectively. Typically, we have observed about 1 three-fold coincidence per two minutes. Our measurements, described in subsequent paragraphs, thus took several hours to accumulate hundreds of coincidences allowing to estimate the results with reasonably small uncertainties (assuming Poissonian distribution of the signal). To compensate for long-term power fluctuations, mainly due to the laser and coupling efficiency fluctuations, we have been swapping in two-minute intervals between two regimes during each measurement: The first regime consists of adjusting the mutual temporal delays between the three photons to be zero and making thus the photons interfere. In the second regime, the temporal overlap between the photons was deliberately detuned so that the photons do not interfere and the observed coincidence rate can be used for normalization.

Our three-photon source does not generate a perfect pure Fock $|1\rangle$ state at each of its outputs. The photon number statistics at the output ports is a product of SPDC and coherent state statistics. These imperfections cause the three-fold coincidences to be observed when, e.g., two photons were present at one of the inputs while another input was in a vacuum state. We call these instances accidental coincidence detections and their rate can be estimated using the theoretical framework reported in Ref. \cite{travnicek}. Knowing the typical single-photon detection rates from the SPDC process and from the attenuated fundamental beam, one can estimate the rate of accidental coincidences and subtract them from the overall detected coincidence rate. This procedure allows to describe the performance of the router independently on the imperfections of the source.

%\section{Testing}
\emph{Testing} --- Once the setup for the quantum router has been constructed and adjusted, we have performed a series of tests to verify that the device works properly, as described by Eq.~(\ref{eq:final}). These tests were performed in three steps, each dedicated to verify one particular property of the operation.

\begin{figure}
\includegraphics[scale=1]{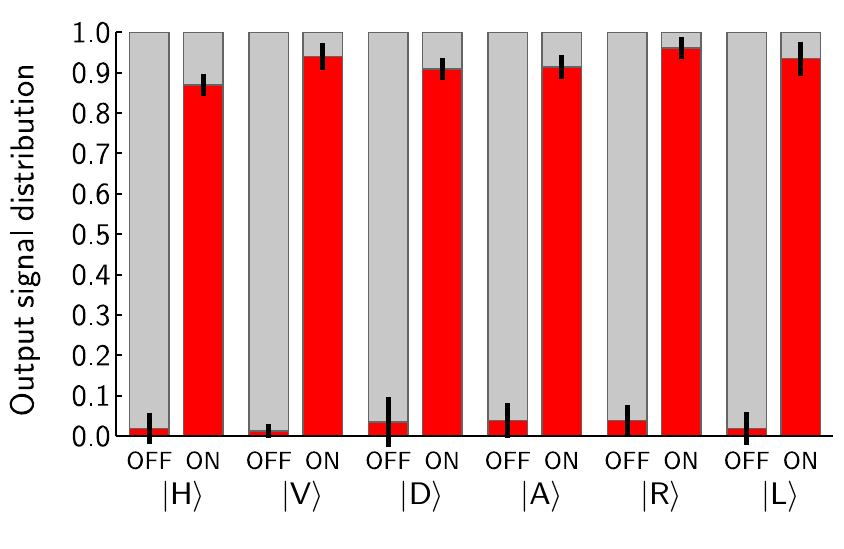}
\caption{\label{fig:ON_OFF}Probability of observing the signal photon leaving the router by the first (lightgrey upper portion of the bar) or second (red lower segment of the bar) output port. The horizontal axis labels indicate the state of the control qubits (ON and OFF) and the state of the signal photon. Black segments centered at the top of each red bar depict the uncertainties of probability estimation. Presented probabilities are corrected by noise subtraction.}
\end{figure}
In the first step, we have verified that depending on the phase shift $\varphi$ the signal is routed to the first or second output respectively. We denote the first control qubit state $|\psi_c\rangle = \frac{1}{\sqrt{2}}\left(|H\rangle + |V\rangle\right)$ ($\varphi = 0$) to be the logical state $|0\rangle$, i.e., the OFF state. Similarly, the first control qubit state $|\psi_c\rangle = \frac{1}{\sqrt{2}}\left(|H\rangle - |V\rangle\right)$ ($\varphi = \pi$) corresponds to the logical state $|1\rangle$, i.e., the ON state. Using this notation together with the Eq. (\ref{eq:final}), one can easily determine that the signal shall leave the device by the first output port, when the control qubits are in the OFF state. In contrary, the signal shall be routed exclusively into the second output port, when the control qubits read ON. For the purposes of this testing stage, we have measured the rate of 3-fold coincidences CC1 and CC2 for both ON and OFF control states and for the signal photon being in one of the following six typical states: horizontal $|H\rangle$, vertical $|V\rangle$, diagonal $|D\rangle$, anti-diagonal $|A\rangle$ linear polarization and right- $|R\rangle$ and left-handed $|L\rangle$ circular polarization. In Fig. \ref{fig:ON_OFF} we depict probabilities of observing the six states of the signal photon in the first and second  output port as a function of the control states ON and OFF (after subtracting the accidental coincidences). Our measurement certifies that the router directs the signal photon to the designated output port with a typical contrast above 20:1 (minimal corrected contrast was 11:1) based on the setting of the control qubits. Tabularized data as well as the raw data without corrections on imperfect three-photon source are presented in the Supplementary Material. The first testing procedure verifies the capability of our device to route the signal correctly depending on the state of the control qubits. 

\begin{figure}
\includegraphics[scale=1]{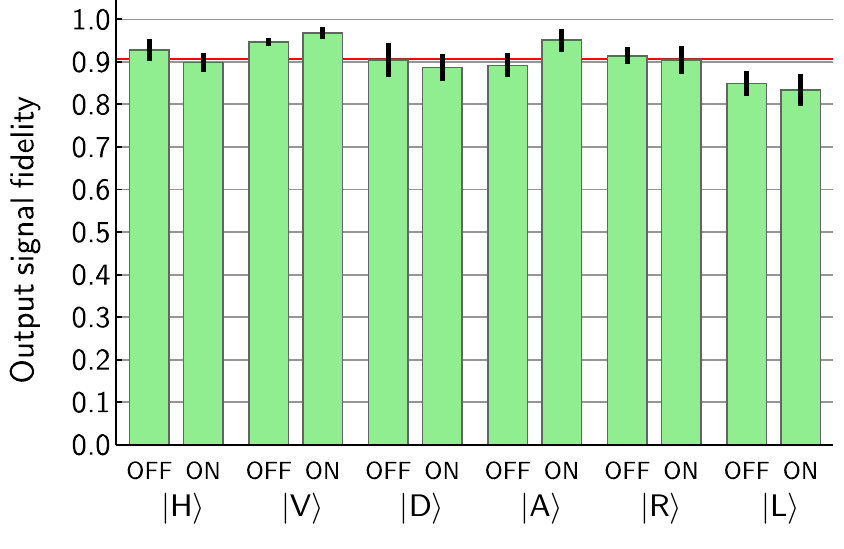}
\caption{\label{fig:fidel}Output signal state fidelities measured for combinations of six input states and control qubit states ON and OFF. Height of green bars correspond to fidelities, black segments centered at the top  at each green bar mark the uncertainties of estimating the fidelities. Red line represents mean value 0.907. Presented fidelities are corrected by noise subtraction.}
\end{figure}

At the second stage we test if the signal state remains undisturbed by measuring output state fidelity for all the combinations of the six input signal states and both control states ON and OFF. The output state fidelity $F = \langle \psi_s| \hat{\rho}_s |\psi_s\rangle$ indicates the overlap between the inserted state $|\psi_s\rangle$ and the generally mixed output signal state $\hat{\rho}_s$. Experimentally, fidelity is obtained by subjecting the output signal photons to polarization projection onto the input signal state and to the orthogonal state. The ratio of coincidence detection rates under these two projections, denoted $R$, gives the fidelity $F = \frac{R}{1+R}$. Fig.~\ref{fig:fidel} shows the observed fidelities after subtracting the accidental coincidences. An average output state fidelity was found to be 0.907 $\pm$ 0.038. For observed values and raw data see Supplementary Material.
At this point, we can certify that the router correctly redirects the signal photon and also quite reliably maintains its state. 

\begin{figure}
\includegraphics[scale=1]{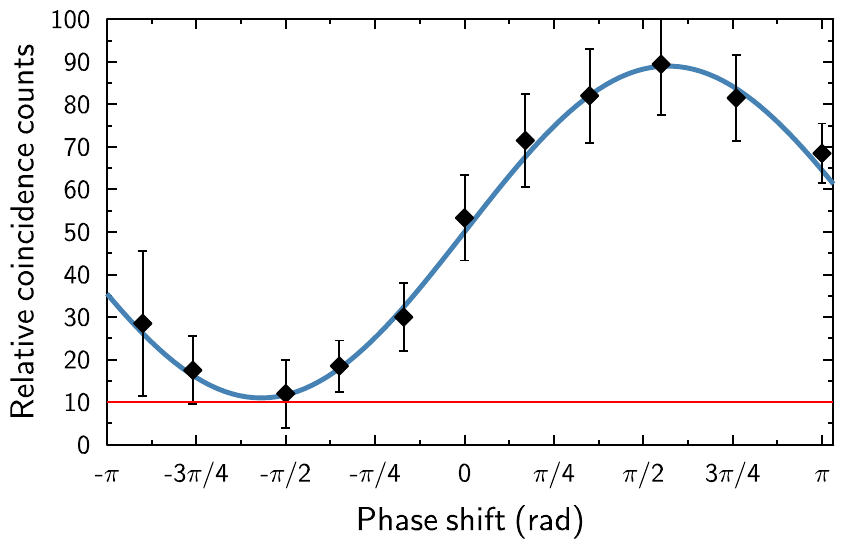}
\caption{\label{fig:coherence} Relative coincidence counts CC1 measured for various phase shifts introduced by the tilt of beam displacer BD4. Black points represent the measured data, blue line is the theoretical fit by a harmonic function and the red line shows the level of accidental coincidences as explained in the text.}
\end{figure}

The last test is to verify the capability of the router to route the signal photon coherently into a superposition of output ports. To check this aspect of the router, we have selected horizontally polarized input signal state $|\psi_s\rangle = |H\rangle$ and set the control qubits to $|\psi_c\rangle = \frac{1}{\sqrt{2}}\left(|H\rangle + i|V\rangle\right)$ which is a balanced superposition between the ON and OFF states. The router setup has been slightly modified to allow us to perform this test. Namely the mirror M1 and beam displacer BD3 have been removed. Projection onto diagonally polarized state was set in the output port S1$_\mathrm{OUT}$. With this modifications of the setup, the signal photon is coherently routed by means of the PPG into two spatial modes which are subsequently overlapped on BD4 (see Fig.~\ref{fig:setup}). By tilting this beam divider, we can introduce an arbitrary phase shift between the two interfering paths and thus observe interference fringes in detected coincidences CC1 behind a polarizer set to project onto diagonally polarized state. The coherence of the routing is thus translated into visibility of these interference fringes. We present our data in Fig.~\ref{fig:coherence} demonstrating that once accidental coincidences are subtracted the visibility reaches $97.7\%\pm0.3\%$. Visibility was calculated using the amplitude of a fitted harmonic function. Tabularized data are provided in the Supplementary Material.

%\section{Conclusions}
\emph{Conclusions} --- In this paper, we have presented our implementation of a linear-optical quantum router. We have tested various aspects of the router to verify correctness of its operation. We have established that the router is able to send the signal state to the designated output port based on the state of the program qubits. The average output signal fidelity is well above the universal cloning threshold of 5/6. 
We have also demonstrated the coherence of routing between the two output ports for a single input state. The coherence of routing should be maintained independently of the input state. In our setup this is ensured by the symmetry of the setup which maintains high fidelity of the output states.

For the purposes of this proof-of-principle experiment, we have operated the router in its basic regime with success probability of 1/16. 
This means that successful router operation was triggered by detecting control qubits in horizontal polarization states.
Note, however, that by adding a classical fiber switch the success rate can be improved to 1/8. In this case such classical switch would simply cross the output fibers when control qubits are both found to be vertically polarized.
With the help of a feed-forward correction the probability of success can even reach 1/4. 
This correction implements the polarization transformation $V \rightarrow -V$ in the signal mode S1 and S2 when the associated control qubit is detected in vertical polarization state (see Ref.~\cite{mikova}). This represents a significant improvement in comparison with previously proposed similar devices.

%\section*{Acknowledgement}
Authors acknowledge
financial support by the Czech Science Foundation under the project No. 17-10003S. 
The authors also acknowledge the project No. CZ.02.1.01/0.0/0.0/16$\_$019/0000754 of the Ministry of Education, Youth and
Sports of the Czech Republic financing the infrastructure of their workplace.

\clearpage
\section*{Supplementary material}

\subsection{Three-photon source}

\begin{figure}[!h!]
\includegraphics[scale=1]{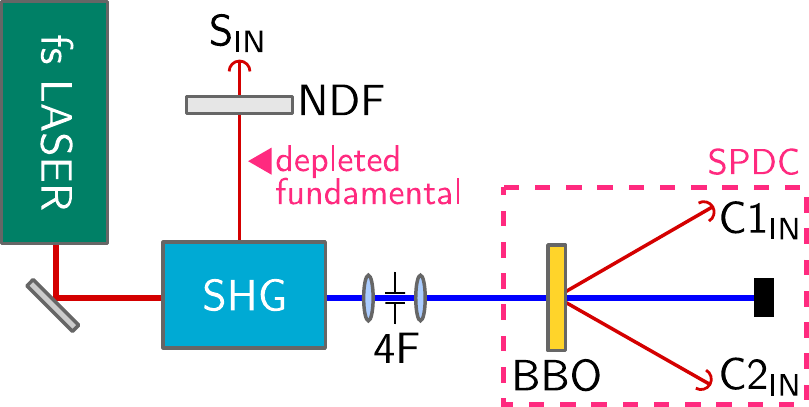}
\caption{Scheme of the three photon source. See description of components in the text.
\label{fig:3psource} }
\end{figure}

In this experiment, we used a typical three-photon source based on spontaneous parametric down-conversion (SPDC) and attenuated coherent state (see Fig. \ref{fig:3psource}). 
We use femtosecond laser system Mira (Coherent) to generate pulses with repetition rate \SI{80}{\mega\hertz}, \SI{800}{\milli\watt} mean power, central wavelength \SI{826}{\nano\meter} and spectral width \SI{10}{\nano\meter} (FWHM). These pulses are frequency doubled in the process of collinear second harmonics generation (SHG). The upconverted light beam in separated on a dichroic mirror. The depleted fundamental mode is attenuated by neutral density filter (NDF) to single-photon level (approximately 0.00125 photons per pulse). The generated second harmonics with central wavelength of \SI{413}{\nano\meter} is filtered spectrally by a pinhole in 4F system. Remaining \SI{100}{\milli\watt} of mean optical power pump nonlinear crystal BBO to produce photon pairs in the Type-I process of spontaneous parametric down-conversion (SPDC). Approximate rate of photon pairs is 2~000 per second. All three optical modes -- attenuated fundamental used as a signal  (S$_{\rm IN}$) and down-conversion used as two controls (C1$_{\rm IN}$,C2$_{\rm IN}$) -- are spectrally filtered by narrow-band filters with \SI{3}{\nano\meter} FWHM. Subsequently the modes are coupled into single-mode optical fibers leading to three optical inputs of the main experimental setup -- linear-optical quantum router.

The main problem of every three photon source of this type is noise caused by multiphoton contributions. In all three modes there is a nonzero probability of having generated two or more photons per laser pulse. And if the spatial modes are mixed on beam splitters as in our router than the multiphoton events can cause false three fold coincidence. These false coincidences have lower probability than the right ones but still they can affect the measurement results. Their rate can be easily found out by sequential blocking of each mode, see analysis in Ref.~\cite{travnicek}. 

\subsection{Measurement procedures and obtained data}

Typical rate of three fold coincidence counts (two controls and one of the signal outputs) was 1--2 per minute. This rate depends on the polarization projection on the signal outputs. To have low errors we have typically accumulated the data for 300 minutes for each setting of the router. Typical probability of accidental coincidences caused by multiple photons was 20\%. Due to the polarization projection the effective rate of noisy coincidences was ten times lower then right ones. 

\begin{figure}
\includegraphics[scale=1]{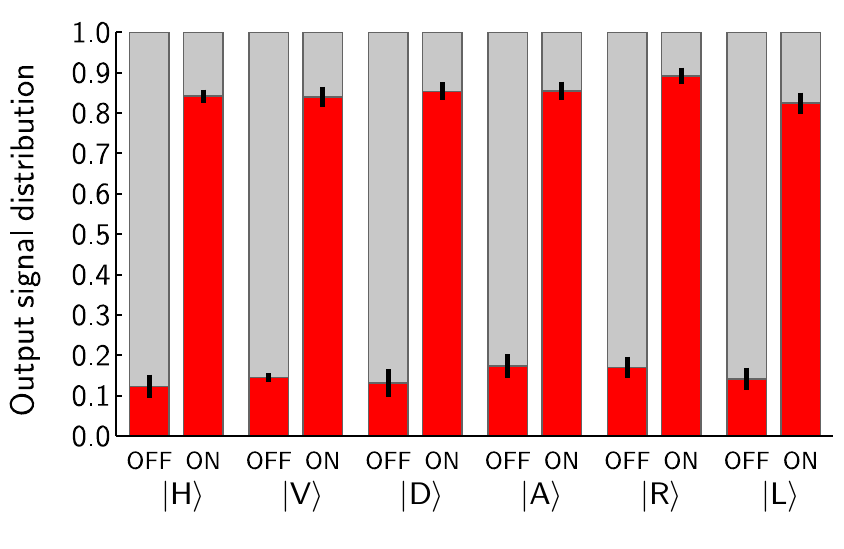}
\caption{\label{fig:ON_OFFraw}Probability of observing the signal photon leaving the router by the first (lightgrey upper portion of the bar) or second (red lower segment of the bar) output port. The horizontal axis labels indicate the state of the control qubits (ON and OFF) and the state of the signal photon. Black segments centered at the top of each red bar depict the uncertainties of probability estimation. Presented probabilities are not corrected by noise subtraction.}
\end{figure}

We present raw data without correction on false three fold coincidences and the corrected data for routing probabilities in the second output port in Table \ref{TAB:result1}. Uncorrected routing probabilities are visualized in Fig. \ref{fig:ON_OFFraw}. 
Mean contrast of the routing in first output mode without correction is $(5.7\pm0.9):1$, with correction then $(15.7\pm4.6):1$. In the second output port we obtain raw contrast $(5.8\pm0.6):1$ and corrected one $(41.8\pm19.7):1$.
\begin{table}[!t!]
\caption{Probability $P_2$ of observing the signal photon leaving the router by the second output port. Probability of observing the signal photon in first output is complement to unity, $P_1 = 1- P_2$. $P_{\mathrm{C}2}$ denotes probability with correction on accidental coincidences. \label{TAB:result1}}
\begin{tabular}{ll|cc|cc}
\hline \hline
signal      & control & $P_2$ & $\sigma P_2$ & $P_{\mathrm{C}2}$ & $\sigma P_{\mathrm{C}2}$ \\ 
\hline 
$|H\rangle$ & OFF &\; 0.123 \; &\; 0.029 \; &\; 0.019 \;&\; 0.039 \; \\
            & ON  & 0.827 & 0.024 & 0.939 & 0.032 \\ \hline
$|V\rangle$ & OFF & 0.145 & 0.011 & 0.012 & 0.017 \\
            & ON  & 0.840 & 0.025 & 0.940 & 0.033 \\ \hline
$|D\rangle$ & OFF & 0.131 & 0.035 & 0.035 & 0.061 \\
            & ON  & 0.854 & 0.022 & 0.909 & 0.028 \\ \hline
$|A\rangle$ & OFF & 0.174 & 0.029 & 0.039 & 0.043 \\
            & ON  & 0.855 & 0.023 & 0.914 & 0.029 \\ \hline
$|R\rangle$ & OFF & 0.170 & 0.026 & 0.039 & 0.039 \\
            & ON  & 0.892 & 0.021 & 0.961 & 0.027 \\ \hline
$|L\rangle$ & OFF & 0.141 & 0.028 & 0.019 & 0.040 \\
            & ON  & 0.825 & 0.026 & 0.935 & 0.042 \\ \hline \hline
\end{tabular}
\end{table}

\begin{figure}
\includegraphics[scale=1]{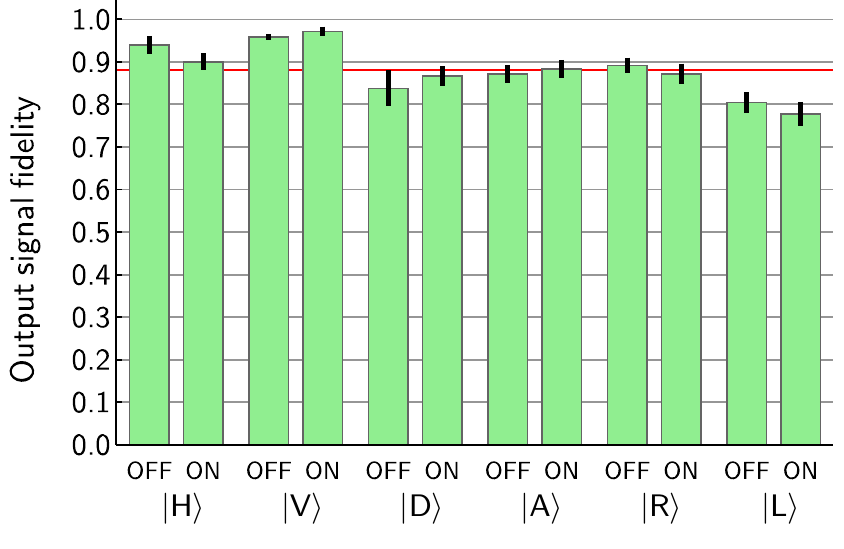}
\caption{\label{fig:fidelraw} Output signal state fidelities measured for combinations of six input states and control qubit states ON and OFF. Height of green bars correspond to fidelities, black segments centered at the top  at each green bar mark the uncertainties of estimating the fidelities. Red line represents mean value 0.881. Presented fidelities are not corrected by noise subtraction.}
\end{figure}

In the table \ref{TAB:result2}, we show measured uncorrected and corrected output state fidelities on the first (control OFF) and the second (control ON) output port. The mean values for both cases are in the last row of the table. The uncorrected fidelities are also shown in Fig. \ref{fig:fidelraw}.
\begin{table}
\caption{Output signal state fidelities measured for combinations of six input states and control qubit states OFF (fidelity measured on the first output) and ON (fidelity measured on the second output). $F_C$ denotes fidelity with correction on accidental coincidences.
\label{TAB:result2}}
\begin{tabular}{ll|cc|cc} 
signal \; & control & $F$ & $\sigma F$ & $F_{\mathrm{C}}$ & $\sigma F_{\mathrm{C}}$ \\ 
\hline \hline
$|H\rangle$ & OFF &\; 0.940 \;&\; 0.021 \;&\; 0.928 \;&\; 0.026 \;\\
            & ON  & 0.900 & 0.020 & 0.899 & 0.022 \\ \hline
$|V\rangle$ & OFF & 0.959 & 0.007 & 0.947 & 0.009 \\
            & ON  & 0.972 & 0.011 & 0.968 & 0.013 \\ \hline
$|D\rangle$ & OFF & 0.838 & 0.042 & 0.905 & 0.040 \\
            & ON  & 0.867 & 0.023 & 0.887 & 0.031 \\ \hline
$|A\rangle$ & OFF & 0.871 & 0.021 & 0.892 & 0.028 \\
            & ON  & 0.883 & 0.022 & 0.951 & 0.027 \\ \hline
$|R\rangle$ & OFF & 0.892 & 0.018 & 0.914 & 0.020 \\
            & ON  & 0.872 & 0.024 & 0.905 & 0.033 \\ \hline
$|L\rangle$ & OFF & 0.805 & 0.024 & 0.849 & 0.029 \\
            & ON  & 0.778 & 0.028 & 0.834 & 0.037 \\ \hline \hline
mean		& &     0.881 & 0.055 & 0.907 & 0.038
\end{tabular}
\end{table}

The coherence between signal outputs OUT1 and OUT2 was tested without mirror M1 and beam displacer BD3 only for one input polarization. The relative phase shift between different paths was tuned by the tilt of the last beam displacer BD4. For each phase we accumulated coincidence typically 50 minutes, see table Tab \ref{tab:coh}. Due to the robust construction the phase in the interferometer remains stable for several hours. Without subtraction accidental coincidences we obtain visibility of interference fringe 76\%. After correction on noise (with value 10.1) we obtain visibility $(97.7\pm0.3)\%$ (calculated from fitted sinus function).
\begin{table}
\caption{Relative values of coincidence counts measured to test coherence between signal outputs. \label{tab:coh}}
\begin{tabular}{ccc}
phase & relative & error \\ 
$[\mathrm{rad}]$ & coincidences & \\ \hline
-0.9  & 28.5 & 17 \\
-0.76 &	17.5 &	8 \\
-0.5  & 12.0 &	8 \\
-0.35 & 18.5 &  6 \\
-0.17 & 30.0 &	8 \\
 0.00 & 53.3 & 10 \\
 0.17 & 71.5 & 11 \\
 0.35 & 82.0 & 11 \\
 0.55 & 89.5 & 12 \\
 0.76 & 81.5 & 10 \\
 1.00 & 68.5 &  7
\end{tabular}
\end{table}

\end{document}